\documentclass[mathpazo]{cicp}
\usepackage{graphicx}
\begin{document}
\title{Double Exchange Model in Triangular Lattice Studied by Truncated Polynomial Expansion Method}
\author[Zhang G. P.]{Gui-ping Zhang\affil{1}\comma\corrauth}
\address{\affilnum{1}\ Department of Physics, Renmin University of China, Beijing 100872, China}
\emails{{\tt zhanggp96@ruc.edu.cn} (G.-P.~Zhang)}
\begin{abstract}
The low temperature properties of double exchange model in
triangular lattice are investigated via truncated polynomial
expansion method (TPEM), which reduces the computational complexity
and enables parallel computation. We found that for the half-filling
case a stable 120$^{\circ}$ spin configuration phase occurs owing to
the frustration of triangular lattice and is further stabilized by
antiferromagnetic (AF) superexchange interaction , while a
transition between a stable ferromagnetic (FM) phase and a unique
flux phase with small finite-size effect is induced by AF
superexchange interaction for the quarter-filling case.
\end{abstract}

 \maketitle

\section{Introduction}
Doped manganite has become one of the most important strongly
correlated systems, since colossal magnetoresistance (CMR) effect
was discovered in 1990s\cite{CMR1,CMR2}. CMR is referred to the
resistivity of material change orders of magnitude under external
magnetic field and it may have a potential application in computer
technology or even spintronics. There are rich phase diagrams and
many ordered phase\cite{phasediagram}, rising from delicate
interaction between electron, spin and orbit degree of freedoms.
Further it is found that the phase separation\cite{phaseseparation}
(PS) may be crucial to CMR effect.

Double-exchange model, as a starting point to study manganite,
describes the Hund interaction between itinerant electron and
localized spin of Mn atoms and is expressed by
\begin{equation}
H_{DE}=-t
\sum_{<ij>,\alpha}(C_{i,\alpha}^{\dag}C_{j,\alpha}+h_{.}c_{.})
-J_{H}\sum_{i,\alpha,\beta}C_{i,\alpha}^{\dag}
\sigma_{\alpha\beta}C_{i,\beta}\cdot S_{i},
\end{equation}
where the nearest-neighbor hopping integral $t$ is adopted as the
energy unit, $J_H$ is the Hund interaction strength,
$C_{i,\alpha}^{\dag}$ ($C_{i,\alpha}$) creates (annihilates) one
electron at site $i$ with spin $\alpha$, $<ij>$ stands for the
nearest neighbors of lattice site, and $\sigma_{\alpha\beta}$ is the
Pauli matrix. The localized spin $S_i$ at site $i$ is assumed as $1$
here. In addition, antiferromagnetic (AF) superexchang is crucial to
stabilize AF phase of some underdopped narrow-band manganite, and
this interaction is expressed by $H_{AF}=J_{AF}\sum_{<ij>}S_{i}\cdot
S_{j}$. So the total hamiltonian is $H=H_{DE}+H_{AF}$. In this
model, localized spins is treated as a classic field $\phi$ and
electron degree of freedom can be integrated for any given localized
spin configuration. The partition function is expressed by
$Z=\text{Tr}_{\text{c}}\text{Tr}_{\text{F}}(e^{-\beta[H(\phi)-\mu
n_{e}]})= \text{Tr}_{\text{c}}e^{-S_{\text{eff}}(\phi)}$, and the
effective action is $S_{\text{eff}}(\phi)=-\sum_{\nu} \ln
(1+e^{-\beta(\epsilon_{\nu}-\mu)})+\beta E(\phi)$. Here
$\epsilon_{\nu}$ is the $\nu$-th eigenvalue of one-electron sector's
Hamiltionian matrix $H(\phi)$, $E(\phi)$ is the interaction energy
between spins, $\beta$ is the inverse temperature, $\mu$ is the
chemical potential, and $n_{e}$ is the electron number operator,
respectively. The fluctuation of localized spins is suitable for
Monte Carlo (MC) simulation, and the partition sum is replaced by
stochastic samplings with the Boltzmann weight
$e^{-S_{\text{eff}}(\phi)}/Z$. With MC simulation, an intrinsic PS
between high electron density and low electron density has been
reproduced in doped maganites\cite{mcs1}.

Despite the success of reproducing PS, the above method is suffered
from finite-size effect, since the computational complexity of exact
diagonalizing (DIAG) $H(\phi)$ scales as $O(N^4)$, with $N$ being
the system size. In order to overcome the above restriction,
Furukawa and Motome \cite{PEM1,PEM2} proposed Chebyshev polynomials
expansion method(PEM) and the computational complexity became
$O(MN^3)$ at a finite cutoff $M$. In 2004 they further reduced the
computational complexity to $O(N)$ via truncated polynomial
expansion method (TPEM)\cite{TPEM1,TPEM2}. The system
size\cite{TPEMappl1} is extended to $40 \times 40$, compared with
$10 \times 10$ via DIAG\cite{mcs1}. Not only the computational
complexity is greatly reduced, but also parallel computation is
capable under PEM and TPEM, which would increase the computational
speed very much. Finally, from the viewpoint of physical properties,
one-body quantity such as energy and electron density, but also
dynamical quantity or two-body quantity such as conductance are able
to be investigated under PEM and TPEM, and there have been several
examples of application to large scale
system\cite{TPEMappl1,TPEMappl2,TPEMappl3,TPEMappl4}.

Up to date, most studies focus on the double-exchange model in one-
and two-dimensional square lattice. In this paper, we present our
results on this model in two dimensional triangular lattice. One
famous layered triangular lattice is the superconductor
$Na_{0.35}CoO_{2}\cdot 1.35H_{2}O$ discovered in 2003
\cite{trilattice}. Though triangular lattice structure is not found
in maganite yet, it is still meaningful to study the interplay
between electron, spin and lattice for double exchange model in two
dimensional triangular lattice. With the frustration, there is
120$^{\circ}$ spin configuration in low temperature for the
half-filling case, and this phase becomes more stable by AF
superexchange interaction; but for the quarter-filling case, we
found a stable long-range ferromagnetic (FM) ordered phase turns
into a unique flux phase induced by AF superexchange interaction.
This paper is arranged as follows: we briefly describe PEM and TPEM
(more details please refer to Ref. \cite{PEM1,PEM2,TPEM1,TPEM2}) and
give some benchmark in Sec 2, in Sec 3 show results in triangular
lattice, and finally we summarize in Sec 4.

\section{The method and benchmark}
The core of Chebyshev polynomial expansion method (PEM) is that the
mean value of physical quantity is expressed by the sum of the
moment multiplied with physical quantity expansion coefficients, and
it converges to exact value guaranteed by rapidly decaying expansion
coefficients. It is important that the accuracy is controlled only
by the cutoff $M$, the thresholds $\epsilon_{p}$ and
$\epsilon_{tr}$, not by the system size and/or the chemical
potential. It is further verified that the error does not accumulate
with Monte Carlo sweep. It is advantageous that the application to
large-scale system is capable, as the computational speed is greatly
improved due to reduced computational complexity. In following, we
will give a brief description of PEM and TPEM, and some benchmark.

\subsection{Polynomial Expansion Method}
The Chebyshev polynomials $T_m(x)$ are recursively defined by: $T_0
(x)=1$, $T_1 (x)=x$, and $T_m (x)=2 x T_{m-1}(x)- T_{m-2}(x)$, with
$ -1 \le x \le 1$. These polynomials are orthonormal in a form that
$$\int_{-1}^{1}\frac{dx}{\sqrt{1-x^2}} T_m (x) T_{m^{'}} (x)=\alpha_m
\delta_{m m^{'}},$$
where $\alpha_m=\left \{ \begin{array}{r@{\quad,\quad}l} 1& m=0,\\
\frac{1}{2}& m \ne 0. \end{array}\right .$ The density of states (DOS)
can be expressed 
as
$D(\epsilon)=\frac{1}{2\pi
\sqrt{1-\epsilon^{2}}}\sum_{m=0}^{\infty}\mu_{m}T_{m}(\epsilon)$,
with the moment being
$\mu_{m}=\int_{-1}^{1}T_{m}(\epsilon)D(\epsilon)d\epsilon$. Then the
mean value of any physical operator $f$ can be expressed by
\begin{equation}
<f>=\int_{-1}^{1}d\epsilon
D(\epsilon)f(\epsilon)=\sum_{m}\mu_{m}f_{m},
\end{equation}
with expansion coefficients $f_m=\frac{1}{\alpha_m} \int_{-1}^{1}
\frac{dx}{\sqrt{1-x^2}}f(x) T_m (x)$, which decay exponentially for
$m \gg 1$ and ensure the accuracy at finite cutoff $M$\cite{PEM2}.
Here $f(x)$ may be the effective action function
$S(x)=-\ln[1+e^{\beta(x-\mu)}]$ or electron number function
$n(x)=1/[1+e^{\beta(x-\mu)}]$, and $f_m$ is easily calculated.

The
moment $\mu_m=\sum_{\nu=1}^{N_{\text{dim}}}T_{m}(\epsilon_{\nu})=Tr(T_{m}(H))$ can be evaluated under any set of
orthonormal basis $e(k)$,
\begin{equation}
\label{eq:moment}
\mu_{m}=\sum_{k=1}^{N_{\text{dim}}}<e(k)|T_m(H)|e(k)>=\sum_{k=1}^{N_{\text{dim}}}\mu_{m}(k),
\end{equation}
Here we define a vector $v(k,m)=T_m(H)|e(k)>$ and a partial moment
$\mu_{m}(k)=<e(k)|v(k,m)>$. From the definition, it is easy to
calculate the vector $v(k,m)$ recursively and the element $i$ in the
vector $v(k,m)$ is expressed as
\begin{eqnarray}
\label{eq:vector}
v_{i}(k,0)=e_{i}(k),\nonumber \\
v_{i}(k,1)=\sum_{j}H_{ij}v_{j}(k,0),\nonumber \\
v_{i}(k,m)=2 \sum_{j}H_{ij}v_{j}(k,m-1)-v_{i}(k,m-2).
\end{eqnarray}


\subsection{Truncation of Matrix-vector Product and Trace Operator}
For a sparse hamiltonian
matrix $H_{ij}$ with $O(N)$ nonzero elements, 
the multiplication can be confined to nonzero elements
that scale as $O(M^D)$ where $D$
is the system dimension. The computational complexity of $M$-th
moment $\mu_M(k)$ is $O(M^{D+1})$ and 
further reduced to $O(M^{D/2+1})$ if elements less than the threshold $\epsilon_p$ are neglected. A subspace $N_{\epsilon_{p}}(k,m)$ is defined by
$N_{\epsilon_{p}}(k,m)=\bigcup_{m^{'}=0}^{m}\{i\}, |v_{i}(k,m^{'})|
\ge \epsilon_{p}$. The total
error of the Boltzmann weight is $O(M^{D+1}\epsilon_{p})$.

Based on the importance sampling method, during each MC sweep a new
configuration is accepted as the ratio $r=\exp^{-\triangle S_{\text{eff}}}$ is larger than a random
number uniformly distributed between 0 and 1. 
The update of the effective action $\triangle
S_{\text{eff}}=\sum_{mk}[\mu_m^{\text{new}}(k)-\mu_m^{\text{old}}(k)]S_m$
is expressed by
\begin{equation}
\label{eq:effact} \triangle S_{\text{eff}}=\sum_m
S_{m}\sum_{k=1}^{N_{\text{dim}}}\triangle\mu_m(k),
\end{equation}
and the update of the moment is
\begin{equation}
\triangle\mu_m(k)=<e(k)|v^{\text{new}}(k,m)>-<e(k)|
v^{\text{old}}(k,m)>.
\end{equation}


Due to local update of the classical field, only a few elements of
the Hamiltonian matrix $H$ are modulated, 
there are only a
limited amount of $v(k,m)$ to be updated, and a new threshold $\epsilon_{\text{tr}}$ can
be taken to reduce the computational complexity.
The total computational complexity of one MC sweep is $O(M^{D+1}N)$.
As Eq.~\eqref{eq:moment} is shown, each basis is inter-independent
and the calculation of the moment $\mu_{m}$ can be parallelized. 

\subsection{Benchmark for double exchange model}

\begin{figure}[htb]
\centering
\includegraphics[width=8.5cm]{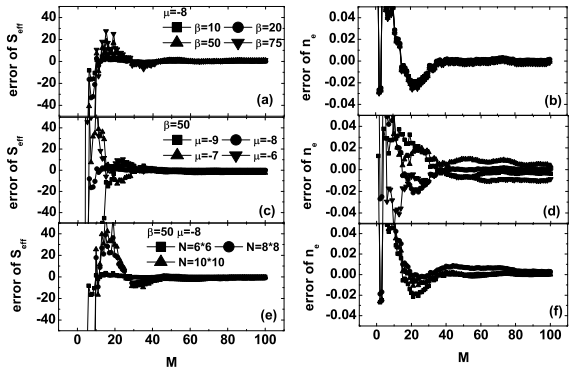}
\caption{  The error of $S_{eff}$ (a, c, e) and $n_e$ (b, d, f) vs
the cutoff $M$ for a given spin configuration in triangular lattice. $J_{H}=8$ and
$J_{AF}=0$. Other parameters are (a) and (b) $N=6 \times 6$, $\mu=-8$; (c)
and (d) $N=6 \times 6$, $\beta=50$ ; and (e) and (f) $\beta=50$,
$\mu=-8$.}\label{figaccuracy}
\end{figure}

Here we present some benchmark, such as the accuracy of the physical quantity,
the number of nonzero elements, and the sweep time.
One basic factor of the algorithm is
the accuracy. As we mention above, the error can be controlled
systematically by $M$, $\epsilon_{\text{p}}$ and
$\epsilon_{\text{tr}}$, and the result is often reliable at finite
$M$. In Fig. \ref{figaccuracy} we show the error of the effective
action $S_{\text{eff}}$ and the electron occupation $n_e$ under a
random spin configuration in two dimensional triangle lattice. As $M$ increases, the
error is smaller and smaller, and the result eventually converges to
the exact value. From Fig. \ref{figaccuracy}(a-b), both the
effective action and the electron density are very accurate for
$20\le M \le 40$, though $M \sim 40$ is required at lower
temperature due to slowly decaying $f_m$. Fortunately, the error
does not depend on other physical parameters, such as the chemical
potential $\mu$ (Fig. \ref{figaccuracy}(c-d)) and the system size
(Fig. \ref{figaccuracy}(e-f)). Moreover, the error of the effective
action does not accumulate during Monte Carlo update process (not
shown here). These facts show that PEM combined with MC simulation
is reliable to study large-scale system.

\begin{figure}[htb]
\centering
\includegraphics[width=7.5cm]{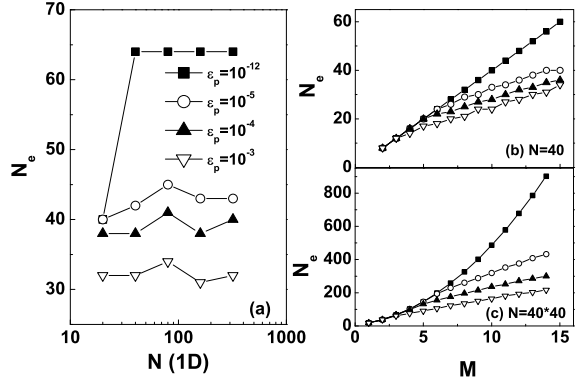}
\caption{ $N_e$ elements in the subspace $N_{\epsilon_{p}}$ vs (a)
the system size $N$ in one dimension, and vs (b, c) the cutoff $M$
in one and two dimension. $\beta=75$, $J_{H}=8$, $\mu=-8$,
$J_{AF}=0$. (a) $M=16$, (b) $N=40$, and (c) $N=40 \times
40$.}\label{figsubspace}
\end{figure}

\begin{figure}[htb]
\centering
\includegraphics[width=8.5cm]{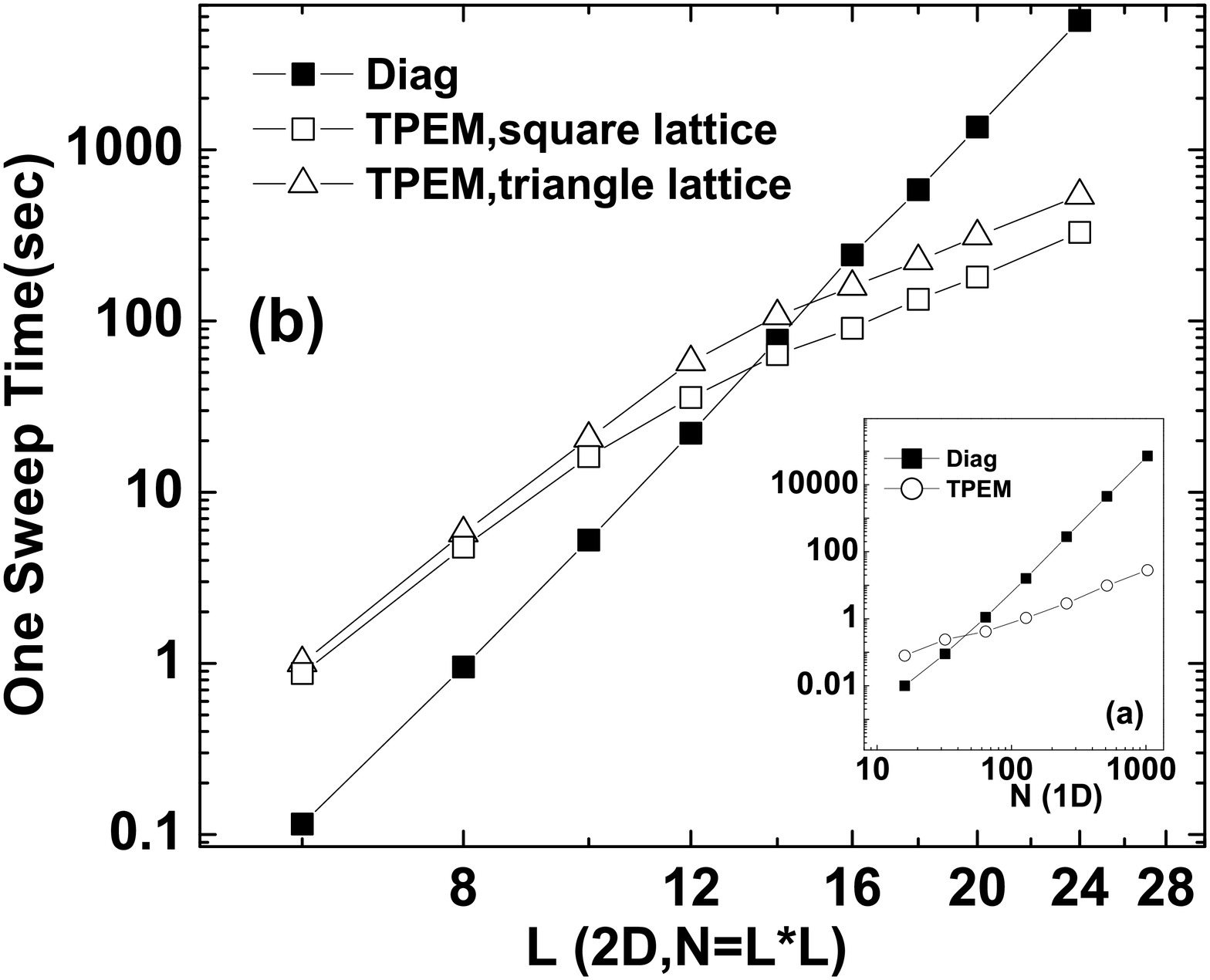}
\caption{ Comparison of one MC sweep CPU time (sec) via TPEM and
DIAG in (a) one and (b) two dimension. $M=30$,
$\epsilon_{pr}=10^{-5}$ and $\epsilon_{tr}=10^{-3}$. }
\label{figCPUtime}
\end{figure}

The other important factor of the algorithm is the computation speed
controlled by computational complexity. For TPEM, the complexity is
usually associated with $N$, the system dimension $D$, $M$,
$\epsilon_{\text{p}}$ and $\epsilon_{\text{tr}}$. In Fig.
\ref{figsubspace}(a), we show how the number of elements in the
subspace $N_{\epsilon_{\text{p}}}$ ($N_e$) change with $N$. While
not surprisingly $N_e$ decreases with higher $\epsilon_{\text{p}}$,
it is interesting that $N_e$ saturates no matter how large $N$ is.
That is why the complexity is linear with $N$ for one MC sweep. In
Fig. \ref{figsubspace}(b-c) the asymptotic behavior of $N_e$ changes
from $O(M^{D})$ to $O(M^{D/2})$ once the truncation
$\epsilon_{\text{p}}$ is taken.
Now we compare the computational speed. Fig. \ref{figCPUtime}
illustrates the CPU time cost by one MC sweep via DIAG and TPEM,
respectively. In one dimension TPEM is prior as $N \ge 64$, and it
is 1000 times faster than DIAG at $N \ge 512$. In two dimension,
TPEM is advantageous as the system is larger than $14\times 14$.
The sweep time is the same in square lattice and triangular
lattice by DIAG , but it is shorter in square lattice than in triangular lattice by TPEM,
as the hamiltonian is sparser owning to electron hopping between less
nearest-neighbors in square lattice. Due to the dimensionality, the scaling behavior of
time with respect to the system size is the same in triangular lattice as
in square lattice. Once parallel computation is realized, TPEM will be more efficient.

\section{Double Exchange Model in Triangular Lattice}
In triangular lattice with the lattice constant $a$, two basic
lattice vectors are $\vec{a}_1=(1,0)a$ and
$\vec{a}_2=(\frac{1}{2},\frac{\sqrt{3}}{2})a$. The reciprocal
lattices are
$\vec{b}_1=\frac{4\pi}{\sqrt{3}a}(\frac{\sqrt{3}}{2},-\frac{1}{2})$
and $\vec{b}_2=\frac{4\pi}{\sqrt{3}a}(0,1)$. For $L \times L$
triangular lattice, the momentum
$\vec{q}=\frac{m}{L}\vec{b}_1+\frac{n}{L}\vec{b}_2$ is shortened as
$(q_1,q_2)$, with $q_1=\frac{m}{L}$, $q_2=\frac{n}{L}$, and $m$, $n$
as integers from $0$ to $L$. Here we mainly focus on the spin
structure factor $S(q)$ defined by
\begin{equation}
S(\vec{q})=\frac{1}{N}\sum_{i,j}<S_i \cdot S_j> e^{i\vec{q} \cdot
\vec{r}_{ij}},
\end{equation}
where $<S_i \cdot S_j>$ is the mean spin-spin correlation,
$\vec{r}_{ij}$ is the displacement between site $i$ and $j$, and
$N=L \times L$.

\begin{figure}[htb]
\centering
\includegraphics[width=8.5cm]{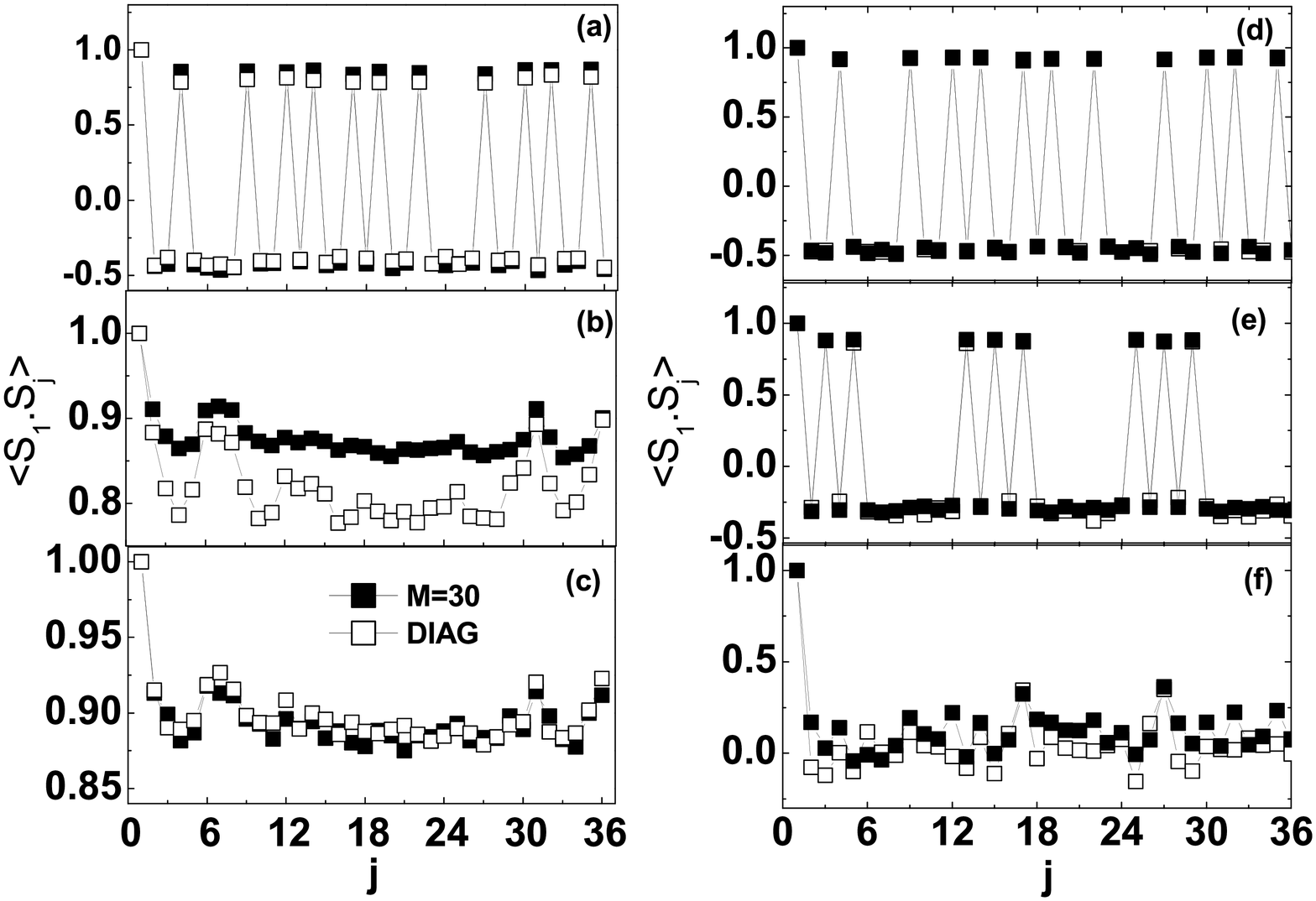}
\caption{ The spin-spin correlation $<S_1 \cdot S_j>$ of $6 \times
6$ triangular lattice. $\beta=75$ and $J_{H}=8$.
$\epsilon_{p}=10^{-5}$ and $\epsilon_{tr}=10^{-3}$. (a) $J_{AF}=0$, $\mu=-6$ and
$<n>=0.9272$, (b) $J_{AF}=0$, $\mu=-8$ and $<n>=0.4512$, (c) $J_{AF}=0$, $\mu=-10$ and
$<n>=0.1944$, (d) $J_{AF}=0.1$, $\mu=-6$ and
$<n>=0.9146$, (e) $J_{AF}=0.1$, $\mu=-8$ and $<n>=0.5$, (f) $J_{AF}=0.1$, $\mu=-10$ and
$<n>=0.2101$.}\label{figspincorrelation}
\end{figure}

In order to understand the interplay between electron, spin and lattice, we
study the spin-spin correlation for different filling case.
First we will check the accuracy of the mean spin-spin correlation.
The maximal monte carlo step is 10000, and the physical quantity
is evaluated every 20 steps after first 2000 warmup steps.
The parameters are $M=30$, $\epsilon_{p}=10^{-5}$ and
$\epsilon_{tr}=10^{-3}$. Due to periodic boundary condition (PBC),
we only show $<S_1 \cdot S_j>$ with $1\le j\le N$ in Fig
\ref{figspincorrelation}. For half-filling case (Fig.
\ref{figspincorrelation}(a, d)) and low-filling case (Fig.
\ref{figspincorrelation}(c, f)) regions, the accuracy are enough. But
for quarter-filling case (Fig. \ref{figspincorrelation}(b)), the
accuracy is not good, and higher $M$ is required.
For the spin flux phase, $M=30$ is enough for the accuracy as Fig. 4(e) shows, and the accuracy was verified
in $6 \times 6$ to $12 \times 12$ triangular lattice.
Therefore it is not necessary to use large $M$ as the basic physical phenomenon still remains.

\subsection{Half-filling case}

For half-filling case one lattice has one electron and electron hops
between adjacent lattices will reduce the total energy. According to
Pauli exclusion rule two electrons on one lattice site should have
opposite spins, hence the nearest-neighbor localized spins tend to
be antiparallel, but there exists frustration effect induced by
triangular lattice. Therefore 120$^{\circ}$ spin configuration
occurs, where three localized spins of each triangle are coplaner
with angle between any two being 120$^{\circ}$. Using the method in
Ref. \cite{spin-flux}, the coplane is verified by the fact that
$(S_i \times S_j)\cdot S_k$ statistically equals to zero, where site
$i$, $j$ and $k$ are located in one triangle; and the angle between
any two being 120$^{\circ}$ is verified by the spin-spin correlation
of adjacent sites being $-0.48$, very close to the limit $-0.5$.
Regarding to $S(\vec{q})$, two peaks are located at $(\frac{1}{3},
\frac{1}{3})$ and $(\frac{2}{3}, \frac{2}{3})$, respectively, and
$S(\vec{q})/N=0.408$ (Fig. \ref{figspinstructure}(a)). This phase is
stabilized by AF superexchange interaction with
$S(\vec{q})/N=0.464$(Fig. \ref{figspinstructure}(d)).

\subsection{Low- and Mediate-filling case}

For low- and mediate-filling case, there is less than one electron
on each lattice site and adjacent spins tend to be parallel. Hence
the system is in FM phase and the peak $S(\vec{0,0})/N=0.821$ is
close to the maximum value 1 (Fig. \ref{figspinstructure}(c)). But
in the presence of superexchange interaction, the system is
paramagnetic (Fig. \ref{figspinstructure}(f)) because of the
competition between electron-mediated FM correlation and AF
superexchange interaction. The spin structure factor is almost flat
at all vector $\vec{q}$ and has no peak at all. But for near zero
filling case, 120$^{\circ}$ spin configuration will be induced by AF
interaction with the interplay between spins and lattice.

\subsection{Quarter-filling case}

For quarter-filling case in one- and two-dimensional square lattice
with PBC, the FM phase is not stable, that is positive spin-spin
correlation at short distance and negative at long distance. This
phenomenon has been observed long before (Ref. \cite{mcs1} and the
Reference inside). Open boundary condition (OBC) or other kind of
boundary condition is chosen to stabilize FM phase{\cite{mcs1}. In
two-dimensional triangular lattice with PBC, the effect of lattice
on FM phase is significant. Each site has 6 nearest-neighbors, and
their spins tend to be parallel. It is found that electron-mediated
FM phase is stablized even at long distance and
$S(0,0)/N=0.899$(Fig. \ref{figspinstructure}(b)). So the FM phase is
strengthened by triangular lattice, compared with square lattice.

Now we consider the effect on AF superexchange interaction on
spin-spin correlation. A spin-flux phase{\cite{TPEMappl1,spin-flux}}
occurs in the square lattice, where four localized spins within each
square lie (anti)clockwise at the same plane. In the triangular
lattice, as $J_{AF}$ increases from 0 to $0.4$ (in fact $J_{AF}$
does not exceed 0.1 in material), the system is in FM phase at first
and then turns into a specific flux phase and eventually evolves
into 120$^{\circ}$ spin configuration phase. For the specific flux
phase corresponding to the mediate AF superexchange interaction,
$(S_i \times S_j)\cdot S_k$ with the sites $i$, $j$ and $k$
belonging to one triangle statistically equals to $0.7$, and the
spin-spin correlation of adjacent sites is $-0.3$. So this phase is
different from both FM phase and 120$^{\circ}$ configuration phase.
The peak $S(\vec{q})$ is located at vectors $(\frac{1}{2}, 0)$, $(0,
\frac{1}{2})$ and $(\frac{1}{2}, \frac{1}{2})$, respectively, and
$S(\vec{q})/N \sim 0.29$ very close to its maximum value 1/3 (Fig.
\ref{figspinstructure}(e)). We further extend the system size from
$6 \times 6$ to $12 \times 12$, and find that the peaks position
does not change and the normalized peak value $S(\vec{q})/N$
converges to 0.285 (Fig. \ref{figpeak}). So finite-size effect is
small and this flux phase is very stable.

We systematically investigate the behavior of the spin-spin correlation and spin structure
factor at high temperature. 
The peak value of spin structure factor decreases with the temperature, and eventually at beta=10 the system is paramagnetic as $J_{AF}$ ranges from 0 to 0.1. So there is no phase transition at high temperature and the system is still subjected to the dimensionality of system.

\begin{figure}[htb]
\centering
\includegraphics[width=8.5cm]{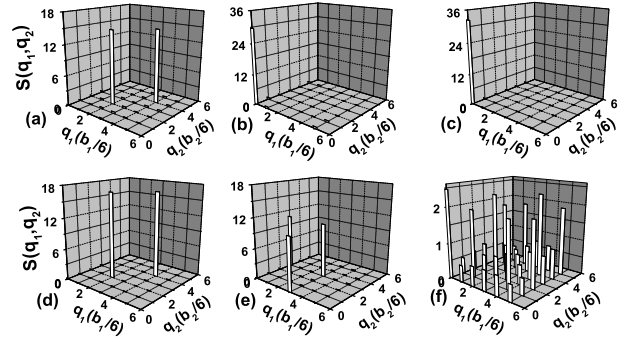}
\caption{ The spin structure factor $S(q_1,q_2)$ of $6 \times 6$
triangular lattice. $\beta=75$, and $J_{H}=8$. (a) $J_{AF}=0$, $\mu=-6$ and
$<n>=0.9272$, (b) $J_{AF}=0$, $\mu=-8$ and $<n>=0.4512$, (c)
$J_{AF}=0$, $\mu=-10$ and $<n>=0.1944$, (d) $J_{AF}=0.1$, $\mu=-6$
and $<n>=0.9146$, (e) $J_{AF}=0.1$, $\mu=-8$ and $<n>=0.5$, and (f)
$J_{AF}=0.1$, $\mu=-10$ and $<n>=0.2101$. }\label{figspinstructure}
\end{figure}

\begin{figure}[htb]
\centering
\includegraphics[width=6cm]{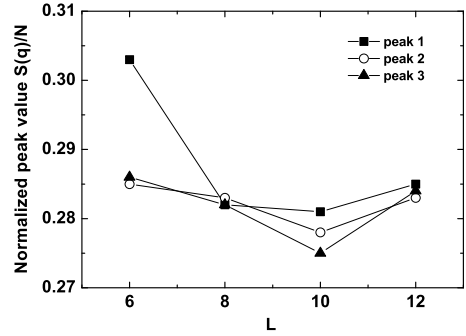}
\caption{The peak $S(\vec{q})$ vs the size $L$ in two dimensional
triangular lattices. $\beta=75$, $J_{H}=8$, $J_{AF}=0.1$, $\mu=-8$ and
$<n>=0.5$. } \label{figpeak}
\end{figure}

\section{Conclusion}

Double exchange model with antiferromagnetic spin-spin superexchange
interaction has been studied for two dimensional triangular lattice
with the truncated polynomial expansion method. For the half-filling
case we obtained 120$^{\circ}$ spin configuration at low
temperature, and it is further stabilized by superexchange
ineraction. For the quarter-filling case, we found that the FM phase
is quite stable, and superexchange interaction results in a unique
spin-flux configuration with very small finite-size effect.

\section*{Acknowledgements} The author would like to thank Prof. X-Q Wang
for proposing this interesting project and many insightful discussions,
and thank Dr Q-L Zhang for discussing Monte Carlo simulation of double exchange model. This work was
supported by the NFSC grants under the numbers 10425417 and 10674142.



\begin{thebibliography}{99}

\bibitem{CMR1} R. von Helmot, J. Wecker, B. Holzapfel,
L. Schultz and K. Samwer, Phys. Rev. Lett., 71 (1993), 2331.

\bibitem{CMR2} S. Jin, T. H. Tiefel, M. McCormakc,
R. A. Fastnacht, R. Ramesh, and L. H. Chen, Science, 264 (1994),
413.

\bibitem{phasediagram} A. Urushibara, Y. Moritomo, T. Arima, A.
Asamitsu, G. Kido, and Y. Tokura , Phys. Rev. B, 51 (1995), 14103.

\bibitem{phaseseparation} Y.-D. Chuang, A. D. Gromko,
D. S. Dessau, T. Kimura, and Y. Tokura, Science, 292 (2001), 1509.

\bibitem{mcs1} S. Yunoki, J. Hu, A. L. Malvezzi, A. Moreo, N. Furukawa, and E.
Dagotto, Phys. Rev. Lett., 80 (1998), 845.

\bibitem{PEM1} Y. Motome, N. Furukawa, J. Phys. Soc. Jpn., 68 (1998),3853.

\bibitem{PEM2} N. Furukawa, Y. Motome, Comput. Phys. Comm., 142 (2001),
410.

\bibitem{TPEM1} N. Furukawa, Y. Motome, J. Phys. Soc. Jpn., 73 (2004),
1482.

\bibitem{TPEM2} G. Alvarez, C. Sen, N. Furukawa,
Y. Motome, and E. Dagotto, Comput. Phys. Commu., 168 (2005), 32.

\bibitem{TPEMappl1} C. Sen, G. Alvarez, Y. Motome, N. Furukawa, I. A. Sergienko, T. C.
Schulthess, A. Moreo, and E. Dagotto, Phys. Rev. B, 73 (2006),
224430.

\bibitem{TPEMappl2} G. Alvarez, T. C. Schulthess, Phys. Rev. B, 73 (2006),
035117.

\bibitem{TPEMappl3} G. Alvarez, H. Aliaga, C. Sen, and E. Dagotto, Phys. Rev. B, 73 (2006)
, 224426.

\bibitem{TPEMappl4} C. Sen, G. Alvarez, H. Aliaga, and E. Dagotto, Phys. Rev. B, 73 (2006),
224441.

\bibitem{trilattice} T. Takada, H. Sakurai, E. Takayama-Muromachi, F. Izumi, R. A. Dilanian and
and T. Sasaki, Nature, 422 (2003), 53.

\bibitem{b8} J. L. Alonso, L. A. Fern\'{a}ndez, F. Guinea, V. Laliena, and V. Mart\'{i}n-Mayo,
Nucl. Phys. B, 596 (2001), 587.

\bibitem{spin-flux} D. F. Agterberg and S. Yunoki,
Phys. Rev. B, 62 (2000), 13816.
\end{thebibliography}
\end{document}